\documentclass[a4paper,10pt]{article}
\usepackage[utf8]{inputenc}
\usepackage{cite}
\usepackage{graphicx}
\usepackage{amsmath}
\def\lsim{\raise0.3ex\hbox{$\;<$\kern-0.75em\raise-1.1ex\hbox{$\sim\;$}}}
\def\gsim{\raise0.3ex\hbox{$\;>$\kern-0.75em\raise-1.1ex\hbox{$\sim\;$}}}
\title{Update of the flavour-physics constraints in the NMSSM}
\author{Florian Domingo}
\date{{\em Instituto de F\'isica de Cantabria (CSIC-UC), E-39005 Santander, Spain\\
Instituto de F\'isica Te\'orica (UAM/CSIC), Universidad Aut\'onoma de Madrid, Cantoblanco,E-28049 Madrid, Spain}}

\begin{document}

\maketitle
\vspace{-5cm}\rightline{IFT-UAM/CSIC-15-130}\vspace{5cm}
\begin{abstract}
We consider the impact of several flavour-changing observables in the $B$- and the Kaon sectors on the parameter space of the NMSSM, in a minimal flavour violating version of this model. Our purpose consists in updating our previous
results in \cite{Domingo:2007dx} and designing an up-to-date flavour test for the public package \verb|NMSSMTools|. We provide details concerning our implementation of the constraints in a series of brief reviews of the current status
of the considered channels. Finally, we present a few consequences of these flavour constraints for the NMSSM, turning to two specific scenarios: one is characteristic of the MSSM-limit and illustrates the workings of charged-Higgs 
and genuinely supersymmetric contributions to flavour-changing processes; the second focus is a region where a light CP-odd Higgs is present. Strong limits are found whenever an enhancement factor -- large $\tan\beta$, light $H^{\pm}$,
resonant pseudoscalar -- comes into play.
\end{abstract}

\section{Introduction}
Flavour-changing rare decays and oscillation parameters are known as uncircumventable tests of the Standard Model (SM) and its new-physics extensions. In the quark sector of the SM, flavour-violation is induced by the 
non-alignment of the Yukawa matrices, resulting in a Cabbibbo-Kobayashi-Maskawa (CKM) mixing matrix, and conveyed only by charged currents at tree-level. While tensions are occasionally reported -- see e.g.\ 
\cite{Descotes-Genon:2015uva} and references therein for a recent example --, this minimal picture seems globally consistent with the current experimental status of flavour observables \cite{Hurth:2012vp}, such that 
new sources or new mediators of flavour violation are relevantly constrained by these measurements. Yet, a proper confrontation of a new-physics model to such experimental results does not depend exclusively on the 
accuracy of the measurements or of the theoretical predictions in the SM, but also on the magnitude of the effects induced beyond the SM (BSM).

In this paper, we will consider the well-motivated supersymmetric (SUSY) extension of the SM known as the Next-to-Minimal Supersymmetric Standard Model (NMSSM) -- see \cite{NMSSM} for a review --, in a minimal flavour-violating 
version: we will assume that the squark sector is aligned with the mass-states in the quark sector, so that, at tree-level, only charged particles convey flavour-violating effects, which are always proportional to the CKM matrix. 
Our aims consist in updating our previous work in \cite{Domingo:2007dx} to the current status of flavour observables and accordingly designing a tool for a test in the flavour-sector which will be attached to the public package 
\verb|NMSSMTools| \cite{NMSSMTools}. Beyond ours, several projects for the study of flavour observables in the NMSSM, or more generally in SUSY extensions of the SM, have been presented in the literature: see e.g.\ 
\cite{Belanger:2004yn,Degrassi:2007kj,Mahmoudi:2007vz,Rosiek:2010ug,Porod:2014xia}.

Our original work dealing with $B$-physics in the NMSSM \cite{Domingo:2007dx} discussed the processes $BR(\bar{B}\to X_s\gamma)$, $BR(B^0_s\to\mu^+\mu^-)$, $BR(B^+\to\tau^+\nu_{\tau})$ as well as the oscillation parameters 
$\Delta M_{d,s}$. These processes had been implemented in the Fortran code \verb|bsg.f| at (grossly) leading order (LO) in terms of the BSM contributions, using the NLO formalism for the SM and locally correcting it to 
account for NNLO effects in an ad-hoc fashion: in other words, this analysis essentially compiled results of the late $90$'s / early $2000$'s \cite{Buchalla:1995vs,Chetyrkin:1996vx,Ciuchini:1997xe,Ciuchini:1998xy,Degrassi:2000qf,
Gambino:2001ew,Hurth:2003dk,Buras:2002vd}. In doing so, it ignored existing NLO results in the MSSM \cite{Bobeth:1999ww,Bobeth:2001jm,Bobeth:2004jz}, focussed instead, at the loop level, on $\tan\beta$-enhanced Higgs-penguin
contributions \cite{Degrassi:2000qf,Buras:2002vd} and only caught the early developments of the NNLO calculation in the SM \cite{Misiak:2006zs,Becher:2006pu}.

The SM analysis of $BR(\bar{B}\to X_s\gamma)$ at NNLO has been recently updated in \cite{Misiak:2015xwa,Czakon:2015exa}: the corresponding results account for significant progress since \cite{Misiak:2006zs} and shift the SM 
expectation $\sim1\sigma$ upwards, very close to the experimental measurement. Similarly, $BR(B^0_s\to\mu^+\mu^-)$ has been considered up to three-loop order in the SM \cite{Bobeth:2013uxa,Hermann:2013kca,Bobeth:2013tba}, 
shifting the result upwards with respect to the LO. Moreover, LHCb and CMS now provide an actual measurement of this process \cite{CMS:2014xfa}, which tightens the associated constraint significantly with respect to the previous upper limits. 
The SM status of $\bar{B}\to X_sl^+l^-$ has also received some attention lately \cite{Huber:2015sra}. Finally, several other channels -- e.g.\ $B^+\to D^{(*)}\tau^+\nu_{\tau}$, the $b\to s\nu\bar{\nu}$ or the $s\to d\nu\bar{\nu}$
transitions -- have been suggested as complementary probes of new physics.

In addition to these recent developments concerning the SM and experimental status of flavour processes, we note that, as the NLO contributions in supersymmetric extensions of the SM can be extracted from e.g.\ \cite{Bobeth:1999ww,
Bobeth:2001jm,Bobeth:2004jz}, it is scientifically sound to include them into our implementation of the observables, so as to reduce the associated uncertainty in the test. This is particularly true in the case of 
$BR(B^0_s\to\mu^+\mu^-)$, since this process is now measured and no longer simply bounded two orders of magnitude from above. The substancial shift in the SM estimate for $BR(\bar{B}\to X_s\gamma)$ also tightens the constraint on 
BSM effects, so that enhanced precision is relevant.

Our purpose in this paper consists in describing the new implementations of flavour observables within \verb|NMSSMTools|. We will first remind succintly of the formalism employed to account for modified Higgs couplings at large 
$\tan\beta$. We will then review briefly each observable and refer explicitly to the literature that we use in their implementation: in a first step, we shall focus on processes in the $B$ sector before turning to Kaon physics. 
Finally, we will illustrate the workings of the new flavour constraints on the parameter space of the NMSSM, comparing the results of our new implementation with the former ones in a few scenarios, and discussing the relevance of
the new observables which have been included.

\section{\boldmath{$\tan\beta$}-Enhanced corrections to the Higgs-quark couplings}\label{epspar}
The Higgs sector of the NMSSM consists of two doublets $H_u=(H_u^+,H_u^0)^T$ and $H_d=(H_d^0,H_d^-)^T$, as well as a singlet $S$. As in the MSSM, the tree-level couplings to quarks involve $H_u$ and $H_d$ in a Type-II 
2-Higgs-Doublet-Model (2HDM) fashion:
\begin{equation}
 {\cal L}_{\mbox{\tiny NMSSM}}\ni Y_u^f\left[H_u^+\left(V^{\mbox{\tiny CKM}}_{ff'}\right)d_L^{f'}-H_u^0u_L^f\right]u_R^{c\,f}+Y_d^f\left[H_d^-\left(V^{\mbox{\tiny CKM}}_{f'f}\right)^*u_L^{f'}-H_d^0d_L^f\right]d_R^{c\,f}+h.c.
\end{equation}
where the diagonal Yukawa parameters can be written in terms of the tree level quark masses and the Higgs vacuum expectation values (v.e.v.'s) -- defined as $v_u=\left<H_u^0\right>=v\sin\beta$, $v_d=\left<H_d^0\right>=v\cos\beta$, with 
$v\equiv\left(2\sqrt{2}G_F\right)^{-1/2}$ --, $Y_u^f=m_u^f/v_u$, $Y_d^f=m_d^f/v_d$ and $V^{\mbox{\tiny CKM}}$ represents the CKM matrix. $f$, $f'$ refer to the generation index.

Yet, radiative corrections, particularly those driven by the SUSY sector, spoil this Type-II picture and generate effective terms such as -- in the $SU(2)\times U(1)$-conserving approximation:
\begin{equation}
 \delta{\cal L}\ni-Y_u^{f}\delta Y_u^{ff'}\left[H_d^{0\,*}u_L^{f'}+H_d^+\left(V^{\mbox{\tiny CKM}}_{f'f''}\right)d_L^{f''}\right]u_R^{c\,f}-Y_d^{f}\delta Y_d^{ff'}\left[H_u^-\left(V^{\mbox{\tiny CKM}}_{f''f'}\right)^*u_L^{f''}+H_u^{0\,*}d_L^{f'}
 \right]d_R^{c\,f}+h.c.
\end{equation}
While in principle a higher-order concern, such terms may be enhanced for large values of $\tan\beta$, so that a resummation becomes necessary for a consistent evaluation.

Here, as in our original work, \cite{Buras:2002vd} remains our main guide. This paper shows that the corrections to the Higgs-quark couplings driven by supersymmetric loops are well approximated in an effective $SU(2)\times 
U(1)$-conserving theory. Corrections to the down-Yukawa couplings and the associated Higgs-quark vertices are dominated by the loop-induced and $\tan\beta$-enhanced contributions to the $H_u^{\dag} q_L d_R^c$ operator, which, 
in turn, can be encoded as the corrections to the down-type quark mass-matrix $\Delta m_d^{ff'}\simeq Y_d^f\delta Y_d^{ff'}v_u$. Corrections to the up-Yukawa are somewhat more subtle, as $\tan\beta$-enhanced terms do not appear 
at the level of the quark masses, but only at that of some -- e.g.\ charged -- Higgs couplings. There, \cite{Buras:2002vd} shows that the parametrization of \cite{Degrassi:2000qf}, complemented by additional corrections, gives 
competitive numerical results. We will be working within these approximations.

In this framework, all the $\tan\beta$-enhanced Higgs-quark vertices can be encoded in terms of the `apparent' quark masses $\bar{m}_q^f$ and CKM matrix elements $V^{eff}_{ff'}$, as well as a bunch of `$\varepsilon$-parameters'
which parametrize:
\begin{itemize}
 \item corrections to the down-type masses: $\tilde{\varepsilon}_d(f)\tan\beta\equiv\mbox{Re}[\Delta m_d^{ff}/m_d^f]$. Such diagonal contributions are mediated by gluino-sdown\footnote{Here as later on, we employ `sup' and `sdown'
in the sense of `up-type sfermions' and `down-type sfermions' without refering to the first generation exclusively} / neutralino-sdown or chargino-sup loops and can be extracted from Eqs.(2.5) and (A.2) in \cite{Buras:2002vd}.
 \item off-diagonal corrections to the down-type mass matrix: $\varepsilon_Y^{ff'}\tan\beta\equiv\Delta m_d^{ff'}/m_d^fY_t^2V_{tf}V_{tf'}^*$
. These, in our minimal flavour violation approximation, are exclusively mediated by chargino-sup loops. Eqs.(2.5) and (A.2) of \cite{Buras:2002vd} again provide explicit expressions in terms of the supersymmetric spectrum.
 \item for the up-type couplings, $\varepsilon'_u(d)$ is defined as the effective correction to the $H_d^+u_R^cd_L$ vertex (see \cite{Degrassi:2000qf}): ${\cal L}\ni u_R^cY_uV_{ud}[H_u^+-\varepsilon'_u(d)H_d^+]d_L$. It is computed 
according to Eqs.(5.6)-(5.8) of \cite{Buras:2002vd}, i.e.\ including relevant electroweak-gauge effects.
 \item corrections to the CKM matrix elements can be encoded in terms of $\tilde{\varepsilon}_0(f)\equiv\tilde{\varepsilon}_d(b)-Y_t^2\varepsilon_Y^{bf}$ so that:
\begin{displaymath}
V_{ff}=V_{ff}^{eff}\hspace{0.5cm};\hspace{0.5cm} V_{bf}=V_{bf}^{eff}\frac{1+\tilde{\varepsilon}_d(f)\tan\beta}{1+\tilde{\varepsilon}_0(f)\tan\beta}
\hspace{0.5cm};\hspace{0.5cm} V_{fb}=V_{fb}^{eff}\frac{1+\tilde{\varepsilon}_d(f)\tan\beta}{1+\tilde{\varepsilon}_0^*(f)\tan\beta}\hspace{0.5cm}(f<b)
\end{displaymath}
\end{itemize}
The relevant flavour-changing Higgs-quark couplings are then given in Eqs.(3.55)-(3.61) and (5.8) of \cite{Buras:2002vd}.

We note that, in this approach, the couplings of the Goldstone bosons expressed in terms of the effective quark masses and CKM elements are formally identical to the tree-level vertices expressed in terms of the tree-level masses
and CKM matrix, so that the Goldstone bosons do not convey explicit $\tan\beta$-enhanced terms.

Another remark addresses the explicit calculation of the $\varepsilon$-parameters: we neglect the Yukawa couplings of the two first generations and assume degeneracy of the corresponding sfermions. Consequently, the unitarity of
the CKM matrix can be invoked in order to include the contributions of both generations at once as, e.g., $V_{cs}^*V_{cb}+V_{us}^*V_{ub}=-V_{ts}^*V_{tb}$.

\section{Observables in the \boldmath{$B$}-sector}
\subsection{\boldmath{$B\to X_{s(d)}\gamma$}}
As mentioned earlier, the status of $B\to X_s\gamma$ in the SM has substancially evolved since the analyses of \cite{Misiak:2006zs,Becher:2006pu}. The new NNLO SM estimate for $E_0=1.6$~GeV \cite{Misiak:2015xwa,Czakon:2015exa} (where $E_0$ is the cut on the photon energy), 
shifted $\sim1\sigma$ upwards with respect to the older estimate, is indeed very close to the experimental measurement \cite{Amhis:2014hma} (combining results from CLEO, Belle and BABAR):
\begin{equation}
 \left.BR[B\to X_s\gamma]\right|_{E_{\gamma}>E_0}^{\mbox{\tiny SM}}=(3.36\pm0.23)\cdot10^{-4}\hspace{0.5cm};\hspace{0.5cm}\left.BR[B\to X_s\gamma]\right|_{E_{\gamma}>E_0}^{\mbox{\tiny exp.}}=(3.43\pm0.22)\cdot10^{-4}
\end{equation}
Trying to account for this result by tuning the c-quark mass in the NLO formalism, as we did in \cite{Domingo:2007dx}, potentially opens new sources of uncertainty. On the other hand, employing the full NNLO formalism is an 
effort-consumming task of limited interest (in our position), considering that BSM effects will be included at NLO only (see below). We therefore settled for a `middle-way', using the NNLO formalism but encoding the pure SM NNLO effects within 
free parameters which are numerically evaluated by comparison with the numbers provided in \cite{Misiak:2015xwa,Czakon:2015exa}. 

In the NNLO formalism, one can write \cite{Misiak:2006zs}:
\begin{equation}
 \left.BR[\bar{B}\to X_s\gamma]\right|_{E_{\gamma}>E_0}=BR[\bar{B}\to X_c e\bar{\nu}]_{\mbox{\tiny exp}}\left|\frac{V_{ts}^*V_{tb}}{V_{cb}}\right|^2
\frac{6\alpha_{em}}{\pi C}\left[P(E_0)+N(E_0)\right]
\end{equation}
where:
\begin{itemize}
 \item $C$ accounts for the normalization using semi-leptonic decays: a current estimate can be extracted from Eqs.(D.1)-(D.3) of \cite{Czakon:2015exa}.
 \item $P(E_0)$ encodes the perturbative contribution in the $\Delta B=1$ OPE in terms of the Wilson coefficients $C_i^{\mbox{\tiny eff}}$ at the 
low-energy scale $\mu_b\simeq2$~GeV. 
From Eqs.(2.10)-(3.11) in \cite{Misiak:2006zs}:
\begin{multline}
\null\hspace{-0.5cm} P(E_0)=\left|C_7^{(0)\mbox{\tiny eff}}(\mu_b)\right|^2
+\frac{\alpha_S(\mu_b)}{4\pi}2\mbox{Re}\left\{C_7^{(0)\mbox{\tiny eff}\,*}(\mu_b)C_7^{(1)\mbox{\tiny eff}}(\mu_b)+\sum_{1\leq i\leq j\leq8}K_{ij}^{(1)}\ C_i^{(0)\mbox{\tiny eff}\,*}(\mu_b)C_j^{(0)\mbox{\tiny eff}}(\mu_b)\right\}\\
+\left(\frac{\alpha_S(\mu_b)}{4\pi}\right)^2\mbox{Re}\left\{\left|C_7^{(1)\mbox{\tiny eff}}(\mu_b)\right|^2+2\sum_{1\leq i\leq j\leq8}K_{ij}^{(1)}\left[C_i^{(1)\mbox{\tiny eff}\,*}(\mu_b)C_j^{(0)\mbox{\tiny eff}}(\mu_b)+C_i^{(0)\mbox{\tiny eff}\,*}(\mu_b)C_j^{(1)\mbox{\tiny eff}}(\mu_b)\right]\right.\\
\left.\null\hspace{4.7cm}+2C_7^{(0)\mbox{\tiny eff}\,*}(\mu_b)C_7^{(2)\mbox{\tiny eff}}(\mu_b)+2\sum_{1\leq i\leq j\leq8}K_{ij}^{(2)}\ C_i^{(0)\mbox{\tiny eff}\,*}(\mu_b)C_j^{(0)\mbox{\tiny eff}}(\mu_b)\right\}\\
+\frac{\alpha_{em}}{4\pi}2\mbox{Re}\left\{C_7^{(0)\mbox{\tiny eff}\,*}(\mu_b)C_7^{(em)\mbox{\tiny eff}}(\mu_b)\right\}+O(\alpha_S^3,\alpha_S\alpha_{em})\label{PE0}
\end{multline}
Beyond the Wilson coefficients at LO $C_i^{(0)\mbox{\tiny eff}}$, QCD NLO $C_i^{(1)\mbox{\tiny eff}}$, QCD NNLO $C_i^{(2)\mbox{\tiny eff}}$ and QED NLO $C_i^{(em)\mbox{\tiny eff}}$, the NLO coefficients 
$K_{ij}^{(1)}$ play a central role in this expression (for simplicity of notations, we factor out $2$ in the case of $K_{ii}^{(1)}$). They can be extracted from Eqs.(3.2)-(3.13) of \cite{Misiak:2006zs} as well as Eqs.(3.1) and (6.3) 
of \cite{Buras:2002tp} and convey the NLO corrections to the partonic process $b\to s\gamma$ as well as the associated Bremsstrahlung contributions. All the $\Phi_{ij}(\delta,z)$ functions entering $K_{ij}^{(1)}$ are
fitted numerically. The only real difficulty lies in incorporating the third line of Eq.~\ref{PE0}, which contains the NNLO Wilson coefficient $C_7^{(2)\mbox{\tiny eff}}(\mu_b)$ and the NNLO coefficients $K_{ij}^{(2)}$.
However, if we confine to the NLO for BSM contributions, we see that these missing quantities originate purely from the SM and may be parametrized as: 
$P_2^{(2)}(SM)+Q_7^{(2)}(SM)\,C_7^{(0)\mbox{\tiny BSM}}(\mu_0)+Q_8^{(2)}(SM)\,C_8^{(0)\mbox{\tiny BSM}}(\mu_0)$, where $\mu_0\simeq160$~GeV is the matching scale and $C_{7,8}^{(0)\mbox{\tiny BSM}}$ only include the BSM effects. Using 
the numerical input from \cite{Misiak:2015xwa} -- see Eqs.(6) and (10) --, it will be possible to identify the coefficients $P_2^{(2)}(SM)$ and $Q_{7,8}^{(2)}(SM)$.
 \item $N(E_0)$ stands for non-perturbative corrections. From Eqs.(3.9) and (3.14) of \cite{Bauer:1997fe}:
\begin{equation}
 N(E_0)\simeq\sum_{1\leq i\leq j\leq8}\tilde{N}_{ij}\ C_i^{(0)\mbox{\tiny eff}\,*}(\mu_b)C_j^{(0)\mbox{\tiny eff}}(\mu_b)\hspace{0.5cm}
\begin{cases}-6\tilde{N}_{17}=\tilde{N}_{27}\simeq-\frac{1}{9m_c^2}\left[\lambda_2+\frac{1}{m_b}(\frac{\rho_1}{3}-\frac{13}{4}\rho_2)\right]\\
 \tilde{N}_{77}=\frac{1}{2m_b^2}\left[\lambda_1-9\lambda_2-\frac{11\rho_1-27\rho_2}{3m_b}\right]
\end{cases}
\end{equation}
where one can use the input from Appendix D -- Eq.(D.1) -- of \cite{Czakon:2015exa}, with the dictionary: $\lambda_1=-\mu_{\pi}^2$; $\lambda_2=\mu_G^2/3$; $\rho_1=\rho_D^2$; $\rho_2=\rho_{LS}^3/3$. This procedure is aimed at 
parametrizing phenomenologically the non-perturbative effects, the parameters being determined in a fit of the semi-leptonic $B$ decays. \cite{Misiak:2015xwa,Czakon:2015exa} then invoke  \cite{Benzke:2010js} to estimate the 
irreducible uncertainties.
 \item We come to the Wilson coefficients at the low scale. These are connected to the Wilson coefficients at the high scale via the Renormalization Group Equations (RGE). For the LO coefficients, the solution to the RGE's -- provided 
$C_i^{(0)\mbox{\tiny eff}}(\mu_0)=\delta_{i2}$, for $i=1,\cdots,6$ -- can be found in Appendix E of \cite{Gambino:2001ew}. Alternatively, one may directly use \cite{Czakon:2006ss}, which allows to derive the NLO coefficients as well:
\begin{equation}
\begin{cases}
 C_k^{(0)\mbox{\tiny eff}}(\mu_b)=\sum_{i=1}^8m_{kl,i}^{(00)}\eta^{a_i}\,C_l^{(0)\mbox{\tiny eff}}(\mu_0)\\
C_k^{(1)\mbox{\tiny eff}}(\mu_b)=\sum_{i=1}^8\eta^{a_i}\left\{m_{kl,i}^{(00)}\,C_l^{(1)\mbox{\tiny eff}}(\mu_0)+\left[m_{kl,i}^{(10)}\eta+m_{kl,i}^{(11)}\right]C_l^{(0)\mbox{\tiny eff}}(\mu_0)\right\}
\end{cases}\end{equation}
with the `magic numbers' $a_i$, $m_{kl,i}^{(00)}$, $m_{kl,i}^{(10)}$ and $m_{kl,i}^{(11)}$ of Tables~1, 3, 4 in the cited reference. Finally, the QED coefficient can be 
obtained from Eqs.(27), (85) and (86) of \cite{Hurth:2003dk} -- multiplied with a factor $\left(\frac{\alpha_{em}}{4\pi}\right)^{-1}$ -- and 
proceeds originally from the study in Ref.\ \cite{Gambino:2001au}.
\end{itemize}

Having sketchily described the general formalism in the previous lines, we are left with the sole remaining task of defining the Wilson coefficients at the matching scale. From the discussion above, it should be clear that, at the 
considered order, we need only the $C_k^{(0)\mbox{\tiny eff}}(\mu_0)$ and $C_k^{(1)\mbox{\tiny eff}}(\mu_0)$. Still, this improves on the treatment in \cite{Domingo:2007dx} where only 2HDM effects were included at NLO.
\begin{itemize}
 \item The SM LO and NLO coefficients are borrowed from \cite{Chetyrkin:1996vx}, Eqs.(11) and (13), where care must be taken, however, to restore the $\mu_0\neq M_W$ dependence. One may also consider Eqs.(28)-(31) and (35)-(40) of 
\cite{Ciuchini:1997xe}.
\item The additional 2HDM contributions are provided by Eqs.(52)-(64) of \cite{Ciuchini:1997xe}.
\item LO contributions from chargino / stop loops were given in \cite{Ciuchini:1998xy} -- Eqs.(4)-(7) -- but the NLO effects in Eqs.(9)-(27) (of the same reference) are not straightforward. Instead, we prefer to use \cite{Bobeth:1999ww,Bobeth:2004jz}. In 
order to avoid the explicit $\ln\frac{\mu_0}{m_{\tilde{T}}}$ in the NLO coefficient, we take care of defining the LO coefficients at the stop (or scharm/sup) scale directly, then running it down to $\mu_0$ via the RGE's -- and taking 
into account the flavour-dependence in the running, i.e.\ the anomalous dimensions $\frac{14}{23}$ and $\frac{16}{23}$ for five flavours become $\frac{14}{21}$ and $\frac{16}{21}$ for six flavours.
\item Finally, for $\tan\beta$-enhanced two-loop effects at the level of the Higgs-quark couplings, we no longer follow \cite{Degrassi:2000qf}, Eqs.(18)-(19) -- which are phrased in terms of the tree-level, and not of the apparent, 
parameters --, but Eqs.(6.51) and (6.53) of \cite{Buras:2002vd}, the former amounting to $0$ for the $G^{\pm}$ contribution in the $SU(2)\times U(1)$-conserving limit. As before, effective neutral Higgs / bottom quark flavour-changing
loops are included -- see Eq.(6.61) in \cite{Buras:2002vd}.
\end{itemize}

At this point, the implementation at SM NNLO + BSM NLO is almost complete. The only remaining task consists in identifying the NNLO coefficients $P_2^{(2)}(SM)$ and $Q_{7,8}^{(2)}(SM)$ numerically. For this, we take good care of 
employing the input parameters described in Appendix D of \cite{Czakon:2015exa} and turning off the BSM contributions. To recover the branching ratio in \cite{Misiak:2015xwa} -- see Eq.(6) of this reference --, we determine a correction 
$P_2^{(2)}(SM)$ of the order of $5\%$ of the total $P(E_0)$. Then, linearizing Eq.~\ref{PE0} in terms of LO BSM coefficients at the matching scale, we find that our implementation should be supplemented with coefficients 
$Q_{7,8}^{(2)}(SM)$ at the permil level in order to recover the numbers appearing in Eq.(10) of \cite{Misiak:2015xwa}. These numbers are of the expected order of magnitude.

Let us finally comment on the error estimate. The SM + CKM + Non-Perturbative uncertainties have been combined in quadrature in Eq.(6) of \cite{Misiak:2015xwa} and we simply double the resulting number $0.23$ in order to obtain 
$2\sigma$ bounds. On top of this SM + CKM + Non-Perturbative error, we add linearly a higher-order uncertainty of $10\%$ on the LO and $30\%$ on the NLO new-physics contributions, each type -- namely 2HDM, SUSY, neutral Higgs --
being added separately in absolute value. To incorporate this uncertainty, we simply use the linearization which has been employed to determine the NNLO parameters just before.

Now let us turn to $BR[\bar{B}\to X_d\gamma]$. $BR[\bar{B}\to X_d\gamma]$ was originally considered in \cite{Hurth:2003dk} at NLO and then, in view of the BABAR measurement \cite{delAmoSanchez:2010ae}, by \cite{Crivellin:2011ba}. 
Finally, \cite{Misiak:2015xwa} extended the analysis to NNLO. Beyond the trivial substitution $s\to d$ in CKM matrix elements, the chief difference with $BR[\bar{B}\to X_s\gamma]$ originates in sizable contributions from the 
partonic process $b\to d\bar{u}u\gamma$ -- since the CKM ratio $\frac{V_{ud}^*V_{ub}}{V_{td}^*V_{tb}}$ is not negligible. The latter can be sampled in several ways -- see e.g.\ \cite{Asatrian:2013raa} --, which provides some 
handle on the associated error estimate. We will be content with the evaluation using constituent quark masses given in Eq.(3.1) of \cite{Asatrian:2013raa}, setting the ratio $\frac{m}{m_b}$ -- with $m$ standing for the mass of 
the light quarks -- in such a way as to recover, in the SM limit, the central value of \cite{Misiak:2015xwa}, Eq.(8):
\begin{equation}
 \left.BR[B\to X_d\gamma]\right|_{E_{\gamma}>E_0}^{\mbox{\tiny SM}}=(1.73^{+0.12}_{-0.22})\cdot10^{-5}
\end{equation}
We can then check the consistency with Eq.(10) of \cite{Misiak:2015xwa} for the new physics contributions.

As before, the SM + CKM + Non-Perturbative uncertainties are taken over from \cite{Misiak:2015xwa}, Eq.(8) -- again we double the error bands to test the observable at the $2\sigma$ level -- and we add linearly the new physics 
uncertainties. On the experimental side, the BABAR measurement \cite{delAmoSanchez:2010ae} has to be extrapolated to the test region, leading to the estimate \cite{Crivellin:2011ba}: 
\begin{equation}
 BR[\bar{B}\to X_d\gamma]_{E_{\gamma}>E_0}^{\mbox{\tiny exp.}}=(1.41\pm0.57)\cdot10^{-5}
\end{equation}

\subsection{\boldmath{$BR[B^0_{s(d)}\to\mu^+\mu^-]$}}
$BR[B^0_s\to\mu^+\mu^-]$ is the observable where the evolution since \cite{Domingo:2007dx} has been the most critical. The experimental status has seen the upper bound $\left.BR[B^0_s\to\mu^+\mu^-]\right|^{\mbox{\tiny exp.}}
<5.8\cdot10^{-8}$ ($95\%$ CL) replaced by an actual measurement at LHCb and CMS \cite{CMS:2014xfa,Amhis:2014hma}: 
\begin{equation}
 \left.BR[B^0_s\to\mu^+\mu^-]\right|^{\mbox{\tiny exp.}}=(3.1\pm0.7)\cdot10^{-9}
\end{equation}
The corresponding value agrees well with the recent SM calculation \cite{Bobeth:2013uxa}: 
\begin{equation}
 \left.BR[B^0_s\to\mu^+\mu^-]\right|^{\mbox{\tiny SM}}=(3.65\pm0.23)\cdot10^{-9}
\end{equation}
It is thus no longer sufficient to consider $\tan\beta$-enhanced effects only, and we therefore design a full test at NLO in the new version of \verb|bsg.f|.

The general formalism remains unchanged and the master formula can be recovered e.g.\ using Eqs.(5.15)-(5.16) of \cite{Bobeth:2001jm}:
\begin{equation}
 BR[B^0_s\to\mu^+\mu^-]=\frac{G_F^2\alpha^2m_{B_s}^5f_{B_s}^2\tau_{B_s}}{64\pi^3\sin^4\theta_W}|V_{tb}V_{ts}^*|^2\sqrt{1-4\frac{m_{\mu}^2}{m_{B_s}^2}}\left\{|c_S|^2+\left|c_P+2\frac{m_{\mu}}{m_{B_s}^2}c_A\right|^2\right\}
\end{equation}
As before, we shall neglect effects from the `mirror operators' -- which are suppressed as $m_s/m_b$ -- and focus on the leading coefficients $c_A$ (pseudovector), $c_S$ (scalar) and $c_P$ (pseudoscalar) of the 
$(\bar{b}s)(\bar{\mu}\mu)$ system. The analysis is simplified by the fact that -- provided the corresponding operators have been suitably normalized -- these semi-leptonic coefficients have a trivial running.

\begin{itemize}
 \item The SM contribution to $BR[B^0_s\to\mu^+\mu^-]$ is known up to three-loop QCD \cite{Hermann:2013kca} and leading QED order \cite{Bobeth:2013tba}. It projects on the pseudovector operator exclusively. We shall use the 
numerical parametrization of \cite{Bobeth:2013uxa}, Eq.(4), to account for it.
 \item Additional 2HDM contributions appear in the form of $Z$-penguins, boxes and neutral-Higgs penguins. \cite{Bobeth:2001jm} provides the corresponding input in Eqs.(3.12), (3.13), (3.32), (3.36)-(3.39), (3.48) and (3.49).
 \item The genuine supersymmetric contributions take the same form and can be found in Eqs.(3.14), (3.16), (3.32), (3.40), (3.42), (3.44), (3.46) of \cite{Bobeth:2001jm}. Instead of using Eq.(3.50)-(3.58) of that same reference for the neutral-Higgs 
penguins, we resort to \cite{Buras:2002vd}, Eqs. (6.35) and (6.36). As in \cite{Domingo:2007dx} -- see also \cite{hiller} --, we replace the squared Higgs mass in the denominator by a Breit-Wigner function, so as to account for 
potentially light Higgs states.
\end{itemize}

The prefactor induces an uncertainty related to CKM, lattice (hadronic form factor) and $B$-width measurement. These are combined in quadrature
at the $2\sigma$ level. In practice, we use: $m_{B_s}=5.36677$ \cite{Agashe:2014kda}, $\tau_{B_s}=(1.607\pm0.010)$~ps \cite{Amhis:2014hma}, 
$f_{B_s}=(226\pm6)$~MeV -- which is an ad-hoc combination of the various results presented in \cite{Aoki:2013ldr} -- and $|V_{tb}V_{ts}^*|=
(41.3\pm1.4)\cdot10^{-3}$ \cite{Ball:2006xx}. Then, an higher-order uncertainty of $2.2\%$ for the SM \cite{Bobeth:2013uxa} and $10\%$ for new physics 
contributions of each type are added linearly.

A similar analysis can be conducted for $BR[B^0_d\to\mu^+\mu^-]$. The experimental measurement \cite{Amhis:2014hma}:
\begin{equation}
 \left.BR[B^0_d\to\mu^+\mu^-]\right|^{\mbox{\tiny exp.}}=(3.9^{+1.6}_{-1.4})\cdot10^{-10}
\end{equation}
combines the LHCb and CMS limits. The formalism is the same as for the $B^0_s$ decay up to the trivial replacement $s\mapsto d$.
In practice, we use the quantities $m_{B_d}=5.27958$ \cite{Agashe:2014kda}, $\tau_{B_d}=(1.520\pm0.004)$~ps \cite{Amhis:2014hma}, 
$f_{B_d}=(188.5\pm5.25)$~MeV -- again an ad-hoc combination of the various results presented in \cite{Aoki:2013ldr} -- and $|V_{tb}V_{td}^*|=
(8.6\pm2.8)\cdot10^{-3}$ \cite{Ball:2006xx}. Due to larger uncertainties, one expects milder limits than in the $B^0_s$ case however.

\subsection{The \boldmath{$b\to sl^+l^-$} transition}
The process $\bar{B}\to X_sl^+l^-$ was not considered in \cite{Domingo:2007dx} but had been added in \verb|bsg.f| later, including only the scalar contributions from
$\tan\beta$-enhanced Higgs penguins. Here we aim at a more complete analysis.

The study in \cite{Huber:2015sra} provides a recent overview of the observables which can be extracted from $\bar{B}\to X_sl^+l^-$. We will confine to the branching fractions in the low -- $[1,6]$~GeV$^2$ -- and high -- 
$\geq14.4$~GeV$^2$ -- $m^2_{l^+l^-}$ ranges. Eqs.(B.33) and (B.36)-(B.38) of the considered paper provide the dependence of these rates on new-physics contributions to the (chromo-)magnetic operators as well as the semi-leptonic operators of the 
vector type. The sole SM evaluation can be extracted from Eqs.(5.13)-(5.15) of \cite{Huber:2015sra}, while the prefactor in Eq.(4.6) of \cite{Huber:2015sra} can be evaluated separately to allow for a different choice of the central values of CKM / non-perturbative contributions: 
we choose to take the latter from \cite{Czakon:2015exa} since the normalization coincides with that of $BR[\bar{B}\to X_s\gamma]$.

The computation of the Wilson coefficients for the (chromo)-magnetic operators -- $C_7^{\mbox{\tiny eff}}$, $C_8^{\mbox{\tiny eff}}$ -- have already been described in connection with $\bar{B}\to X_s\gamma$: we simply run these
coefficients down to the matching scale of \cite{Huber:2015sra}, $\mu_0'=120$~GeV. Moreover, $C_{10}$ coincides with $c_A$ -- discussed in the context of $B_s^0\to\mu^+\mu^-$ -- up to a normalization factor. Only
$C_9$ is thus missing: it can be obtained in \cite{Bobeth:2004jz} -- see Eq.(3.6) and Appendix A of this reference. Although the lepton flavour has very little impact on $C_{9,10}$ -- it intervenes only via the lepton Yukawa couplings in 
subleading terms --, we still distinguish among $C^e_{9,10}$ and $C^{\mu}_{9,10}$.

While this is ignored by \cite{Huber:2015sra}, $\bar{B}\to X_sl^+l^-$ could also be mediated by scalar operators as shown in Eq.(2.5) of \cite{Cornell:2003qt} -- note that the coefficients $C_{Q_{1,2}}$ there coincide, up to a 
normalization factor, with $c_{S,P}$ introduced before. Therefore, we add these contributions accordingly, estimating the integrals over $m^2_{l^+l^-}$ numerically. To account for possibly light Higgs states, the Higgs-penguin 
contributions from SUSY loops are isolated in $c_{S,P}$ and receive denominators of the form $m^2_{l^+l^-}-m_{h_i^0}^2+\imath m_{h_i^0}\Gamma_{h_i^0}$, which are then integrated. Note that the scalar coefficients $c_{S,P}$ depend 
linearly on the lepton mass, so that they matter only in the case of the muonic final state.

Finally, we come to the error estimate: the SM uncertainties (including e.g.\ CKM effects) are extracted from Eqs.(5.13)-(5.15) of \cite{Huber:2015sra}. Linearizing Eqs.(B.33) and (B.36)-(B.38) (of that same reference) in terms of $C^{\mbox{\tiny BSM}}_{7,\cdots,10}$, we associate 
a $10\%$ uncertainty to these new-physics contributions and add it linearly. As the contributions from scalar operators is added `by hand', we use a larger uncertainty of $30\%$. The experimental values relevant for the 
$\bar{B}\to X_sl^+l^-$ transition are extracted from Eqs.(1.1) and (1.2) in \cite{Huber:2015sra}.

The normalized FB asymmetry $\bar{A}_{FB}[\bar{B}\to X_sl^+l^-]$ could also be implemented using the results in \cite{Huber:2015sra}:
\begin{equation}
\bar{A}_{FB}[q^2_{\mbox{\tiny min}},q^2_{\mbox{\tiny Max}}] \equiv\int_{q^2_{\mbox{\tiny min}}}^{q^2_{\mbox{\tiny Max}}}dq^2\int_{-1}^1\mbox{sgn}(z)dz\frac{d^2\Gamma}{dq^2dz}
\Big/\int_{q^2_{\mbox{\tiny min}}}^{q^2_{\mbox{\tiny Max}}}dq^2\int_{-1}^1dz\frac{d^2\Gamma}{dq^2dz}=\frac{3}{4}\frac{{\cal H}_A}{{\cal H}_T+{\cal H}_L}[q^2_{\mbox{\tiny min}},q^2_{\mbox{\tiny Max}}]
\end{equation}
with the quantities ${\cal H}_A$, ${\cal H}_T$ and ${\cal H}_L$ explicited in Appendix B of \cite{Huber:2015sra}. 
Note that the contributions from scalar operators are suppressed as $\left(\frac{m_l}{m_b}\right)^2$ \cite{Cornell:2003qt} and may thus be neglected.
However, the only experimental source (Belle) \cite{Sato:2014pjr} chose a different binning, so that the results cannot be compared.

Beyond the inclusive decay rates, much effort has been mobilized in the study of the $B\to K^{(*)}l^+l^-$ exclusive modes in the last few years. The full angular analysis
of these modes provide two dozen independent observables \cite{Altmannshofer:2008dz}. Tensions with the SM estimates have been reported in some of these channels, however, leading to a substancial literature 
(see e.g.\ \cite{Altmannshofer:2013foa,Descotes-Genon:2013wba,Descotes-Genon:2015uva}).
In this context, we choose to disregard these exclusive modes for the time being, waiting for a clearer understanding of the reported anomalies.

\subsection{The \boldmath{$b\to s\nu\bar{\nu}$} transition}\label{bsnn}
The $b\to s\nu\bar{\nu}$ transition is known to provide theoretically clean channels. While ignored in our original work, we decide to include the three following observables in the new version of the code:
$BR[B\to X_s\nu\bar{\nu}]$, $BR[B\to K\nu\bar{\nu}]$ and $BR[B\to K^*\nu\bar{\nu}]$.

We follow the analysis of \cite{Buras:2013ooa} (section 5.9), updated in \cite{Buras:2014fpa}. The Wilson coefficients are provided at NLO in section 3.2 of \cite{Bobeth:2001jm}. Under our assumption
of minimal flavour violation, with no flavour-changing gluinos or neutralinos, and neglecting the masses of the light quarks, only the coefficient $C_L$ (or $X_L$ in the notations of \cite{Buras:2013ooa}) 
receives contributions in the model. The relation between the branching ratios in the NMSSM and that in the SM thus becomes particularly simple: see Eqs.(229-232) of \cite{Buras:2013ooa}. We employ the updated SM 
evaluations in Eqs.(10), (11) and (23) of \cite{Buras:2014fpa}:
\begin{equation}
 \begin{cases}
  \left.BR[B\to X_s\nu\bar{\nu}]\right|^{\mbox{\tiny SM}}=(2.9\pm0.3)\cdot10^{-5}\\
  \left.BR[B^+\to K^+\nu\bar{\nu}]\right|^{\mbox{\tiny SM}}=(3.98\pm0.47)\cdot10^{-6}\\
  \left.BR[B^0\to K^{*0}\nu\bar{\nu}]\right|^{\mbox{\tiny SM}}=(9.19\pm0.99)\cdot10^{-6}
 \end{cases}
\end{equation}
We also note that the ratio of the $B^+$ / $B^0$ lifetimes controls that of the $B^+\to K^{(*)+}$ / $B^0\to K^{(*)0}$ transitions.

The experimental upper bound on the inclusive branching ratio $BR[B\to X_s\nu\bar{\nu}]$ originates from ALEPH \cite{Barate:2000rc};
those on the exclusive modes $BR[B\to K\nu\bar{\nu}]$ and $BR[B\to K^*\nu\bar{\nu}]$ are controlled by BABAR \cite{Lees:2013kla} and BELLE \cite{Lutz:2013ftz} respectively
(see also the compilation in \cite{Amhis:2014hma}). At $90\%$ CL:
\begin{multline}
 \null\hspace{5cm}\left.BR[B\to X_s\nu\bar{\nu}]\right|^{\mbox{\tiny exp.}}<6.4\cdot10^{-4}\\\begin{cases}\left.BR[B^+\to K^+\nu\bar{\nu}]\right|^{\mbox{\tiny exp.}}<1.6\cdot10^{-5}\\\left.BR[B^0\to K^0\nu\bar{\nu}]\right|^{\mbox{\tiny exp.}}
 <4.9\cdot10^{-5}\end{cases}\hspace{0cm};\hspace{0.5cm}
 \begin{cases}\left.BR[B^+\to K^{*+}\nu\bar{\nu}]\right|^{\mbox{\tiny exp.}}<4\cdot10^{-5}\\\left.BR[B^{0}\to K^{*0}\nu\bar{\nu}]\right|^{\mbox{\tiny exp.}}<5.5\cdot10^{-5}\end{cases}
\end{multline}

Generalizing to the $b\to d\nu\bar{\nu}$ transition is trivial, though not competitive at the moment.

\subsection{Flavour transitions via a charged current}
The central observable in the $b\to u$ transition is $BR[B^+\to\tau^+\nu_{\tau}]$. Here, we perform little modification of the original implementation in \cite{Domingo:2007dx}. In other words, we follow \cite{Akeroyd:2003zr},
where the effects of the $W$ and charged-Higgs exchanges at tree-level, corrected by $\tan\beta$-enhanced supersymmetric loops, appear in Eq.(5-7) of the quoted paper. The uncertainty is assumed dominated by $V_{ub}$ \cite{Agashe:2014kda} and the 
hadronic form factor \cite{Aoki:2013ldr}.

The $b\to c$ transition has focussed some attention in the last few years. We will consider the ratios $R_D\equiv\frac{BR(B^+\to D\tau^+\nu_{\tau})}{BR(B^+\to Dl^+\nu_{l})}$ and 
$R_{D^*}\equiv\frac{BR(B^+\to D^*\tau^+\nu_{\tau})}{BR(B^+\to D^*l^+\nu_{l})}$.

These quantities show a tension between the SM predictions from lattice / HQET $R_D^{\mbox{\tiny SM}}=0.297\pm0.017$, 
$R_{D^*}^{\mbox{\tiny SM}}=0.252\pm0.003$ \cite{Kamenik:2008tj,Fajfer:2012vx,Lees:2012xj} or, more recently, $R_D^{\mbox{\tiny SM}}=0.299\pm0.011$ \cite{Lattice:2015rga},
$R_D^{\mbox{\tiny SM}}=0.300\pm0.008$ \cite{Na:2015kha} and the experimental averages 
$R_D^{\mbox{\tiny exp}}=0.391\pm0.050$, $R_{D^*}^{\mbox{\tiny exp}}=0.322\pm0.022$ \cite{Amhis:2014hma} (HFAG website), which
combine results from BABAR \cite{Lees:2012xj}, LHCb \cite{Aaij:2015yra} and Belle \cite{Huschle:2015rga}. Note that these 
tensions in the $b\to c$ transition are independent from the CKM uncertainty on $V_{cb}$, due to the normalization.

We follow the analysis in \cite{Crivellin:2012ye} which presents the corrections to the observables in a 2HDM context, allowing to account for 
modified Higgs-quark vertices with respect to Type II, such as those induced by $\tan\beta$-enhanced supersymmetric loops: see Eqs.(7-15) in the reference under consideration.
\begin{equation}
 \begin{cases}
  R_D=R_D^{\mbox{\tiny SM}}\left\{1+1.5\mbox{Re}[C_c^R+C_c^L]+|C_c^R+C_c^L|^2\right\}\\
  R_{D^*}=R_{D^*}^{\mbox{\tiny SM}}\left\{1+0.12\mbox{Re}[C_c^R-C_c^L]+0.05|C_c^R-C_c^L|^2\right\}
 \end{cases}
\end{equation}
These contributions are mediated by a charged Higgs and we can easily translate, for the charged-Higgs / quark couplings, the notations of 
\cite{Crivellin:2012ye} to ours (see section \ref{epspar}). We find the following Wilson coefficients:
\begin{equation}
 \begin{cases}
C_c^R=\frac{m_bm_{\tau}}{m_{H^{\pm}}^2}\left\{\frac{1+\tan^2\beta}{1+\tilde{\varepsilon}_0(s)\tan\beta}-1\right\}\\
C_c^L=-\frac{m_cm_{\tau}}{m_{H^{\pm}}^2}\left\{1-\frac{1+\tan^2\beta}{\tan\beta}\left[\varepsilon'_c(b)-\frac{\varepsilon_Y^{23}\tan\beta}{1+\tilde{\varepsilon}_0(s)\tan\beta}\left(\varepsilon'_c(s)-\varepsilon'_c(b)\right)\right]\right\}
 \end{cases}
\end{equation}

We assume a $30\%$ uncertainty on these new-physics coefficients, which we add linearly to the SM uncertainty quoted above. Since these observables
are only marginally compatible with the SM prediction, we do not devise an actual test for them, but simply propose an evaluation.

\subsection{\boldmath{$B_{d,s}^0$ oscillation parameters $\Delta M_{d,s}$}}\label{DMq}
The old version of the code used the formalism of \cite{Buras:2002vd} to encode the SM -- see Eq.(6.7) of that work -- as well as the $\tan\beta$-enhanced 
double-penguin contributions -- Eq.(6.12)-(6.22) of \cite{Buras:2002vd} --, while the one-loop BSM boxes -- see Eq.(6.3) of \cite{Buras:2002vd} -- were taken from Eq.(94)-(98) of 
\cite{Bertolini:1990if}. (For the latter, only the contributions involving charged particles are relevant under
our assumption of minimal flavour violation, i.e.\ the box contributions mediated by gluinos or neutralinos and sdowns vanish.)

In the new version of the code, we fully upgrade the approach to $\Delta M_{d,s}$ to the NLO formalism:
\begin{itemize}
 \item The box contributions from charged Higgs / tops and chargino / squarks are matched onto the relevant base of operators -- see e.g.\ Eq.(2.1) 
in \cite{Buras:2001ra} -- according to the formulae in Appendix A.4 of \cite{Buras:2002vd}. They are run down from the new-physics scale 
-- $m_{H^{\pm}}$ and the squark scale respectively -- down to the matching scale of $166$~GeV via RGE solutions for 6 flavours: see \cite{Buras:2001ra}, 
Appendix C. As before, SM and double-penguin contributions are also included within this formalism.
 \item We follow section 3.1 of \cite{Buras:2001ra} to connect the matching scale to the low-energy matrix elements -- 5-flavour running.
 \item The low-energy physics is described by the so-called `Bag' parameters -- matrix elements of the operators. We rely essentially on the lattice 
calculations of \cite{Becirevic:2001xt}, except for the operator $Q^{VLL}$, which receives the SM contributions and has thus attracted more recent
attention. In this later case, we use the current FLAG average \cite{Aoki:2013ldr} for $\hat{B}_{B_{d,s}}$ -- which coincides with the Bag parameter
up to a rescaling.
\end{itemize}
These ingredients allow to derive a prediction for $\Delta M_{B_{d,s}}$ using the master formula of \cite{Buras:2002vd}, Eqs.(6.6)-(6.8). The 
hadronic form factors $f_{B_{s,d}}$ are taken from \cite{Aoki:2013ldr}, where we combine the various results. For the CKM elements, we continue
to rely on the evaluation from tree-level processes proposed in \cite{Ball:2006xx}. Note that our central value for $\Delta M_s$ in the SM limit 
is somewhat higher than the latest estimates \cite{Lenz:2010gu,Lenz:2011ti}. This is essentially due to the choice for the lattice input: 
\cite{Lenz:2010gu,Lenz:2011ti} have their own averaging, leading to a smaller form factor, while we follow \cite{Aoki:2013ldr}.

We come to the error estimate. The uncertainty associated to the SM contributions to the operator $Q^{VLL}$ is often neglected in the literature.
Eq.(11) of \cite{Lenz:2010gu} shows however that there could be an error of at least a few permil. We therefore associate a $1\%$ uncertainty to
this contribution, which we add linearly to a $30\%$ uncertainty on each type -- charged Higgs Box / SUSY Box / double penguin -- of new-physics
contribution. Then, the uncertainties on the Bag parameters are taken from \cite{Becirevic:2001xt,Aoki:2013ldr} and combined in quadrature at the 
$2\sigma$ level. Finally, we factor out the uncertainties on the CKM \cite{Ball:2006xx} and lattice form factor \cite{Aoki:2013ldr}, adding them
in quadrature at the $2\sigma$ level. Note that the CKM uncertainty dominates the total error on $\Delta M_{d}$ (at the level of $60\%$) and is 
actually of the order of magnitude of the central value, so that it is important not to linearize the associated error.

These results are then confronted to the experimental measurements \cite{Amhis:2014hma}:
\begin{equation}
 \left.\Delta M_{d}\right|^{\mbox{\tiny exp.}}=(0.5055\pm0.0020)~\mbox{ps}^{-1}\hspace{0.5cm};\hspace{0.5cm}\left.\Delta M_{s}\right|^{\mbox{\tiny exp.}}=(17.757\pm0.021)~\mbox{ps}^{-1}
\end{equation}

\section{Observables in the Kaon-sector}

\subsection{The \boldmath{$s\to d\nu\bar{\nu}$} transition}
The physics of Kaons also provides limits on new physics, one example being the $s\to d \nu\bar{\nu}$ transition. We again follow \cite{Buras:2013ooa} (section 5.8), together with the updated SM results of 
\cite{Buras:2015qea}.

The Wilson coefficients mediating the transition \cite{Buras:2004qb} are very similar to those intervening in the $b\to s \nu\bar{\nu}$ transition, except that the interplay of CKM matrix elements is formally different. The 
normalization to $V_{td}V_{ts}^*$, instead of $V_{tb}V_{ts}^*$, gives more weight to the two first generations: for completion, we thus incorporate the effects proportional to the charm Yukawa coupling. Note that we continue 
to neglect the quark masses of the first generation and that tree-level neutralino and gluino couplings do not mediate flavour transitions (per assumption), so that only the $X_L$ (in the notations of \cite{Buras:2013ooa} and 
equivalent to the $C_L$ of section \ref{bsnn}) coefficient is relevant.

The decay of a charged kaon to $\pi^+\nu\bar{\nu}$, as well as that of the neutral $K_L$ to $\pi^0\nu\bar{\nu}$, can then be encoded in terms of this Wilson coefficient: see Eqs.(213-214) of \cite{Buras:2013ooa}, where, however, 
we substitute the more recent SM input of Eqs.(2.2), (2.9), (2.11) of \cite{Buras:2015qea}.

On the experimental side, the process involving charged mesons is constrained by \cite{Artamonov:2008qb} to $BR[K^+\to \pi^+\nu\bar{\nu}]=(17.3_{-10.5}^{+11.5})\cdot10^{11}$; for the neutral mesons, \cite{Ahn:2009gb} provides 
the bound  $BR[K_L\to \pi^0\nu\bar{\nu}]<2.6\cdot10^{-8}$ ($90\%$ CL).

\subsection{\boldmath{$K-\bar{K}$} mixing}
As for the $B$ mesons, one can consider the mixing of $K$ and $\bar{K}$ mesons. Associated quantities are very precisely measured experimentally \cite{Agashe:2014kda}:
\begin{equation}
 \left.\Delta M_K\right|^{\mbox{\tiny exp.}}=(0.5293\pm0.0009)\cdot10^{-2}\mbox{ps}^{-1}\hspace{0.5cm};\hspace{0.5cm}\left|\varepsilon_K\right|^{\mbox{\tiny exp.}}=(2.228\pm0.011)\cdot10^{-3}
\end{equation}
where $\Delta M_K$ stands for the mass-difference between $K_L$ and $K_S$ and $\varepsilon_K$ measures indirect CP-violation in the $K-\bar{K}$ system.

However, the theoretical estimates of these quantities of the $K-\bar{K}$ system suffer from a substancial uncertainty associated to long-distance effects. Several estimates based on representations of large $N$ QCD
had been proposed in the $80^{\mbox{\tiny ies}}$ -- see e.g.\ \cite{Bijnens:1990mz}. Lately, lattice collaborations have been emphasizing the possibility to perform an evaluation in a realistic kinematical configuration 
in the near future: see e.g.\ \cite{Bai:2014cva}. We will follow \cite{Buras:2013raa} (see also discussion and literature therein) in estimating the long-distance contribution to $\Delta M_K$ at $(20\pm10)\%$ of the experimental
value, while \cite{Bailey:2015tba} provides some lattice input for $\varepsilon_K$: we take over the quantity $\xi_0$ from Eq.(74) and the error estimate on $\xi_{LD}$, Eq.(75) (of that reference) -- these values were originally computed in
\cite{Blum:2011ng} and \cite{Christ:2012se}, and Eq.(67) of \cite{Bailey:2015tba} explicits their impact on $\varepsilon_K$.

We now turn to the short-distance contributions in the $K-\bar{K}$ system. The discussion is very similar to the case of the $B-\bar{B}$ mixing in section \ref{DMq}. We follow \cite{Brod:2011ty} for the SM part: this paper performed 
a NNLO evaluation of the charm contribution -- see Eq.(15) in that work --, completing earlier results for the mixed charm-top \cite{Brod:2010mj} and top \cite{Buras:1990fn} contributions. Note that \cite{Buras:2013raa} and \cite{Bailey:2015tba} 
also propose recent evaluations of the quantity $\eta_{cc}$, with slightly lower central value, but we choose to stick to the conservative estimate of \cite{Brod:2011ty}. The master formulae for $\Delta M_K$ and $\varepsilon_K$ 
are provided in Eqs.(18) and (16) of this reference, respectively -- see also Eqs.(XVIII.6-9) of \cite{Buchalla:1995vs}, as well as Eqs.(XII.3-5) for the expression of the functions $S_0$. The kaon mass of $0.4976$~GeV and the 
form factor $f_K=0.1563\pm0.0009$~GeV are taken from \cite{Agashe:2014kda} and \cite{Aoki:2013ldr} respectively.

The inclusion of BSM contributions follows the general NLO formalism of \cite{Buras:2001ra}: see Eqs.(7.24-7.32). Eqs.(3.20-3.38) (of this reference) explicit the running between the matching- and the low-energy scales.
However, we will be using more recent `Bag parameters' for a low-energy scale of $3$~GeV: Table XIII of \cite{Jang:2015sla} compiles several recent lattice calculations, which we put to use. As in the $B-\bar{B}$ case,
the Wilson coefficients account for Higgs / quark and chargino / squark box diagrams as well as Higgs double-penguin contributions and we follow appendix A of \cite{Buras:2002vd}. Yet, we also include effects associated
to the charm Yukawa, as the interplay of CKM elements gives more weight to such terms than for the $B-\bar{B}$ mixing. Finally, we use appendix C of \cite{Buras:2001ra} to run each new-physics contribution from the 
relevant BSM scale (charged-Higgs or squark mass) down to the matching scale of $166$~GeV.

The SM uncertainty, where the uncertainty on the $\eta_{cc}$ parameter and the long distance effects dominate, is added linearly to the uncertainty driven by the bag parameters and a $30\%$ error on higher-order contributions
to the BSM Wilson coefficients.

Note that one can lead a similar analysis for the $D-\bar{D}$ mixing, which we do not consider here, however.

\section{Sampling the impact of the flavour constraints}
Based on the discussion of the previous sections, we design two Fortran subroutines \verb|bsg.f| and \verb|Kphys.f| for the evaluation in the NMSSM of the considered observables in the $B$- and the Kaon-sectors (respectively), as
well as a confrontation to experimental results. These subroutines are then attached to the public tool \verb|NMSSMTools| \cite{NMSSMTools}.

\subsection{Comparing the new and the old codes}
\begin{figure}[!htb]
    \centering
    \includegraphics[width=13cm]{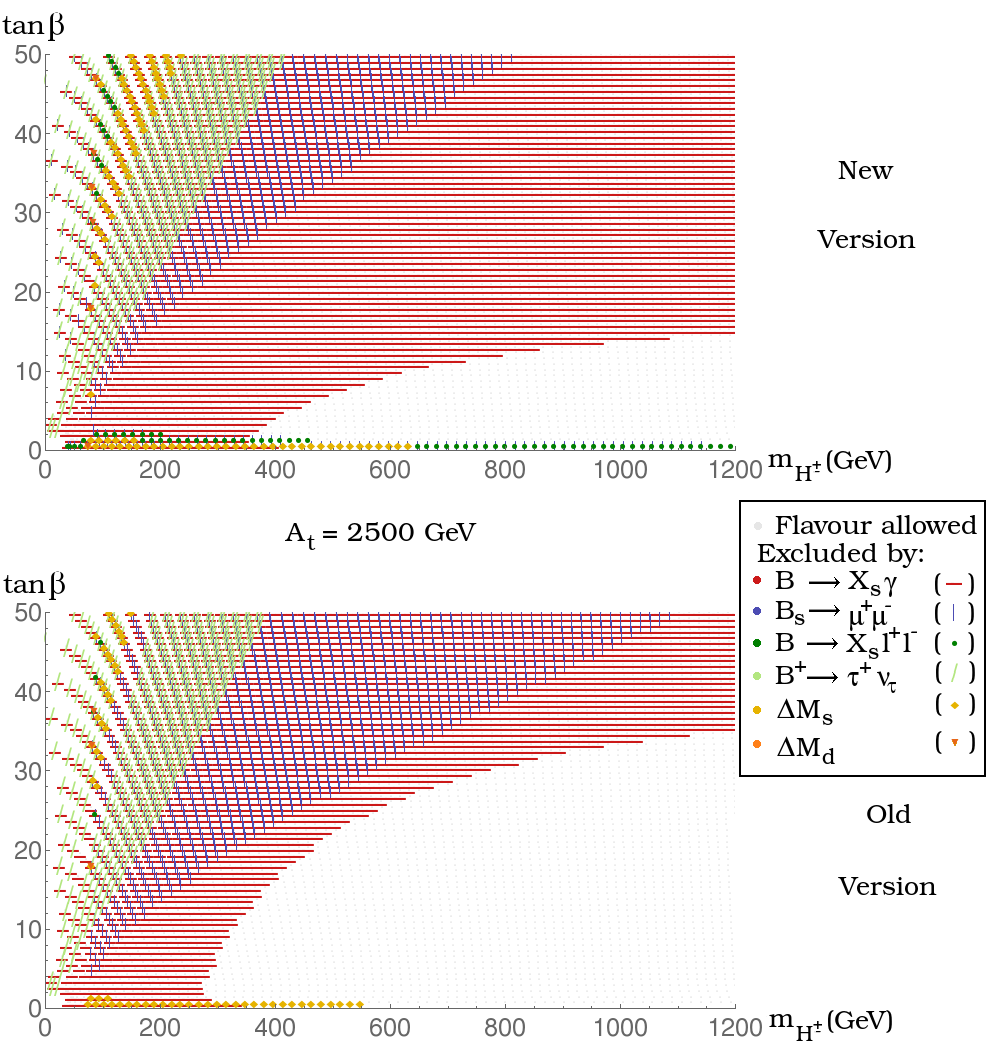}
  \caption{Exclusion contours due to the flavour constraints in the plane $\{m_{H^{\pm}},\tan\beta\}$ for $\lambda=\kappa=2\cdot10^{-4}$, 
$\mu=2M_1=M_2=M_3/5=300$~GeV, $m_{\tilde{F}}=1$~TeV, $A_t=2.5$~TeV, $A_{b,\tau}=-1.5$~TeV, $A_{\kappa}=-500$~GeV. The plot on the first row has been obtained with the new code while the plot of the second row results from
the old version. Points excluded by the various constraints are coloured in red (horizontal lines) -- $\bar{B}\to X_s\gamma$ --, blue (vertical lines) -- $B_s^0\to\mu^+\mu^-$ --,
dark green (circles) -- $\bar{B}\to X_sl^+l^-$ --, light green (oblique lines) -- $B^+\to\tau^+\nu_{\tau}$ --, yellow (diamonds) -- $\Delta M_s$ -- and orange (triangles) -- $\Delta M_d$ --, while remaining gray
points satisfy all these limits.}
  \label{MHC-TB}
\end{figure}

\begin{figure}[!htb]
    \centering
    \includegraphics[width=13cm]{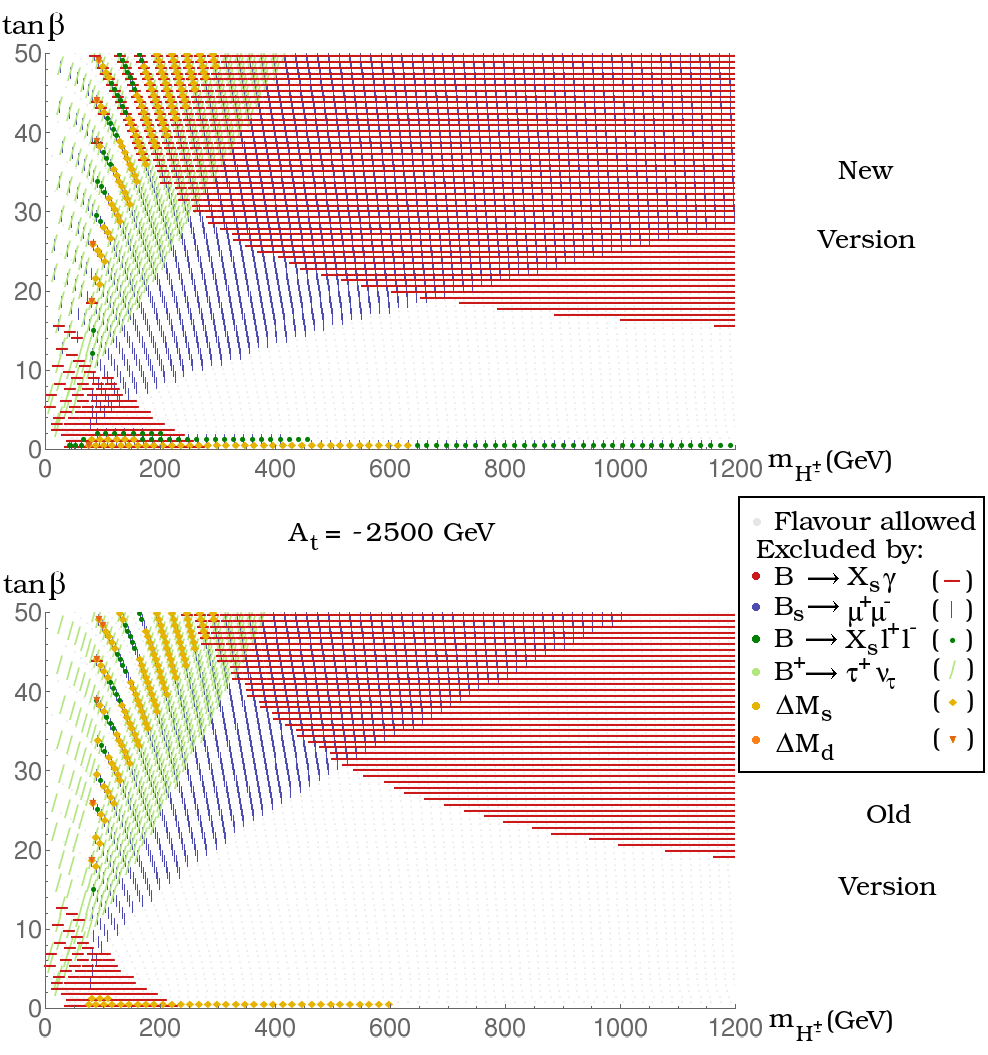}
  \caption{Similar plots and scan as in Fig.~\ref{MHC-TB}, but with $A_t=-2.5$~TeV.}
  \label{MHC-TB_m}
\end{figure}

In order to test the differences between the new and the former implementations of $B$-physics observables in \verb|bsg.f|, we perform a scan over the plane defined by $m_{H^{\pm}}$ -- the charged 
Higgs mass -- and $\tan\beta$ and display the exclusion contours associated with flavour constraints in Figs.~\ref{MHC-TB} and \ref{MHC-TB_m}. The chosen region in the NMSSM parameter space
corresponds to the MSSM-limit, with degenerate sfermions and hierarchical neutralinos. Note that we disregard the phenomenological limits 
from other sectors (e.g.\ Higgs physics, Dark matter, etc.). We consider a large value of the trilinear stop coupling $|A_t|=2.5$~TeV, which is known 
to enhance effects driven by supersymmetric loops, and study separately the two opposite signs -- a negative value of $A_t$, when $\mu>0$, typically triggers
destructive interferences among the SUSY and 2HDM contributions to $\bar{B}\to X_s\gamma$.

The general appearance of the exclusion contours in Figs.\ref{MHC-TB} and \ref{MHC-TB_m} remains qualitatively similar, when comparing the results obtained with the new (plots on the top) and old (plots on the bottom) versions of the 
code\footnote{Note that the old version of the code had been updated to include recent experimental values, so that the differences with the new implementation are fully controlled by the theoretical treatment of the observables.}.
Yet, quantitatively, one witnesses a few deviations:
\begin{itemize}
 \item The limits from $\bar{B}\to X_s\gamma$ are more severe in the new version, which is mostly apparent in Fig.~\ref{MHC-TB} ($A_t>0$): this is not unexpected since the larger SM central value -- closer to the experimental 
measurement -- correspondingly disfavours new physics effects which interfere constructively with the SM contribution (2HDM effects or supersymmetric loops for $\mu,\,A_t>0$). Consequently, the areas with a light charged Higgs
or large $\tan\beta$ receive excessive BSM contributions in view of the experimental measurement and are thus disfavoured. Moreover, note that the full NLO implementation reduces somewhat the error bar associated to higher-order 
new-physics contributions, which also results in tighter bounds for the more recent code. For $A_t<0$, one observes two separate exclusion regions: for low values of $\tan\beta$ and $m_{H^{\pm}}$, the 2HDM contribution is large 
(excessive) while the negative SUSY effect is too small to balance it; on the contrary, with large $\tan\beta$ and heavy $H^{\pm}$, the SUSY contribution dominates and is responsible for the mismatch with the experimental 
measurement. In between, the destructive interplay between the SM and 2HDM effects on one side and the SUSY loops on the other succeeds in keeping $BR[\bar{B}\to X_s\gamma]$ within phenomenologically acceptable values.
\item Limits from $B_s^0\to\mu^+\mu^-$ used to be little sensitive to the sign of $A_t$ in the older implementation. This is no longer true, the reason being that the scalar coefficients $c_{S,P}$ receive new contributions, which 
(had been neglected in the previous version of the code and) may interfere constructively or destructively with the Higgs-penguin effects. This channel appears as the most sensitive one, together with $\bar{B}\to X_s\gamma$,
in the considered scenario. Given the shape of the exclusion regions driven by $\bar{B}\to X_s\gamma$, however, $B_s^0\to\mu^+\mu^-$ seems most relevant for $A_t<0$ (Fig.~\ref{MHC-TB_m}). Expectedly, the limits are tighter for 
large $\tan\beta$, where SUSY contributions are enhanced.
\item Limits from $\bar{B}\to X_sl^+l^-$ differ more significantly between the two implementations -- although they remain subleading. In particular, an excluded region appears at
low $\tan\beta$: it is largely driven by the 2HDM contributions to the semi-leptonic vector coefficients $C_{9,10}$ -- which indeed involve terms in $\tan^{-1}\beta$. On the other hands, the exclusion region at low
$m_{H^{\pm}}$ is largely unchanged: it is associated with the enhancement of the Higgs-penguin contributions for a light Higgs sector.
\item Despite the corrections to the $\tan\beta$-enhanced Higgs/quark vertices, the constraints from $B^+\to\tau^+\nu_{\tau}$, $\Delta M_s$ and
$\Delta M_d$ are little affected by the modernization of the code and remain subleading.
\end{itemize}
We observe that $\bar{B}\to X_s\gamma$ and $B_s^0\to\mu^+\mu^-$ intervene as the determining limits from the flavour sector in the considered scenario: they exclude all the region beyond $\tan\beta\gsim20$.
The low $m_{H^{\pm}}$-region is in tension with most of the observables in the $B$-sector (unsurprisingly), though $\bar{B}\to X_s\gamma$ and $B_s^0\to\mu^+\mu^-$ again appear as the limiting factors
at low-to-moderate $\tan\beta$. Interestingly, $\bar{B}\to X_sl^+l^-$ seems to offer a competitive test for $\tan\beta\lsim2$.

We perform a second test in a region involving a light CP-odd Higgs state with mass below $15$~GeV -- still presuming nothing of the limits from other sectors: note that this is a phenomenologically viable scenario in the NMSSM,
although the limits on unconventional decays of the SM-like Higgs state at $\sim125$~GeV place severe constraints on the properties of the light pseudoscalar. The results are displayed in Fig.~\ref{MA1-TB} -- in terms of the
mass of the pseudoscalar $m_{A_1}$ and $\tan\beta$ -- and confirm the trends that we signaled before:
\begin{figure}[!htb]
    \centering
    \includegraphics[width=13cm]{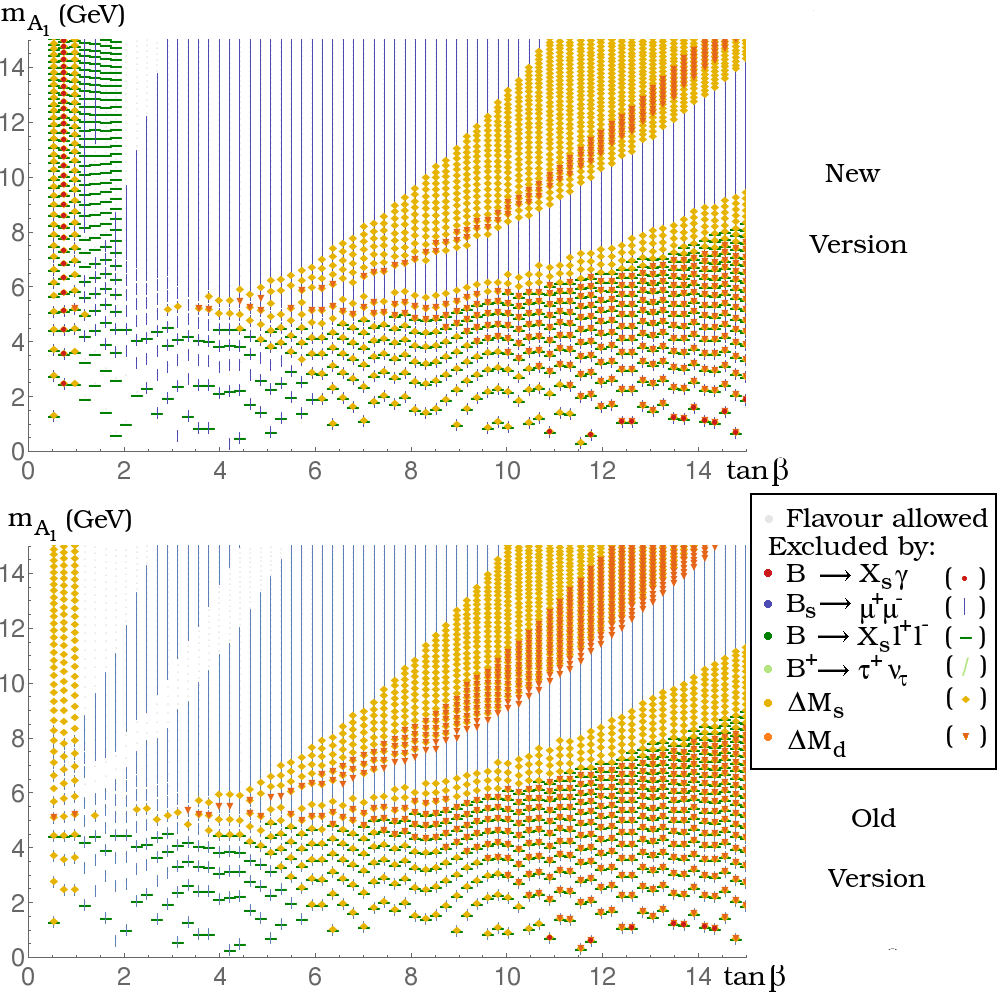}
  \caption{Exclusion contours due to the flavour constraints in the plane $\{\tan\beta,m_{A_1}\}$ for $\lambda=0.45$ $\kappa=0.4$, 
$\mu=2M_1=M_2=M_3/5=300$~GeV, $m_{\tilde{F}}=1$~TeV, $A_{b,\tau}=-1.5$~TeV, $A_t=-2.5$~TeV, $A_{\kappa}=-30$~GeV. Above, the 
results with the new code; below, the results of the old version. The colour code remains the same as before. The symbols are also unchanged, except for $\bar{B}\to X_sl^+l^-$, with horizontal lines, and,
$\bar{B}\to X_s\gamma$, with circles (for reasons of lisibility).}
  \label{MA1-TB}
\end{figure}
\begin{itemize}
 \item Limits from $\bar{B}\to X_s\gamma$ intervene here at low $\tan\beta$ -- where the supersymmetric contributions cannot balance the effect triggered by the charged-Higgs (note that $A_t<0$). A few points are also excluded 
for low $m_{A_1}$ and large $\tan\beta$: these result from the two-loop effect mediated by a neutral Higgs. They prove subleading in the considered region.
\item Limits from $B_s^0\to\mu^+\mu^-$ appear somewhat tighter in the new implementation. In particular, a narrow corridor where the new physics effects reverse the SM contribution is visible in the plot on the bottom of Fig.~\ref{MA1-TB} 
(which corresponds to the older implementation of the limits) -- from $(m_{A_1}\sim6,\tan\beta\sim2)$ to $(m_{A_1}\sim15,\tan\beta\sim5)$; this region is no longer accessible with the more recent code (it is, in fact, shifted to lower 
values of $\tan\beta$).
This channel is the main flavour limit in the considered region, due to the large contribution mediated by an almost on-shell Higgs penguin.
\item Limits from $\bar{B}\to X_sl^+l^-$ intervene in two fashions. One is the exclusion driven by an almost-resonant pseudoscalar and the associated bounds are 
essentially unchanged with respect to the older implementation. Additionally, a new excluded area appears at low $\tan\beta$.
\item Limits from $\Delta M_{d,s}$ are qualitatively unchanged among the two versions, though the bounds associated with $\Delta M_d$ seem somewhat more conservative in 
the new implementation. These constraints remain subleading however, in view of the more efficient $BR(B_s^0\to\mu^+\mu^-)$, and confine to the resonant regime 
-- note e.g.\ the allowed `corridor' where new-physics contributions reverse the SM effect -- or the very-low range $\tan\beta\lsim1$.
\item Limits from $B^+\to\tau^+\nu_{\tau}$ do not intervene here. 
\end{itemize}
$B_s^0\to\mu^+\mu^-$ thus appears as the constraint which is most sensitive to the enhancement-effect related to a near-resonant pseudoscalar. The exclusion effects are most severe for larger $\tan\beta$ as the 
Higgs-penguin is correspondingly enhanced. For $\tan\beta\lsim2$, $\bar{B}\to X_sl^+l^-$ proves a sensitive probe in its new implementation.

Note that, in the two scenarios that we discussed here, the precise limits on the $\{m_{H^{\pm}},\tan\beta\}$ or $\{\tan\beta,m_{A_1}\}$ planes of course depend on the details of the parameters. In particular, the large value of 
$|A_t|$ triggers enhanced SUSY effects, resulting in severe bounds on the considered planes. We thus warn the reader against over-interpreting the impression that only corners of the parameter space of the NMSSM are in a position 
to satisfy $B$-constraints at $95\%$ CL, as Figs.~\ref{MHC-TB}, \ref{MHC-TB_m} and \ref{MA1-TB} might lead one to believe.
\begin{figure}[!htb]
    \centering
    \includegraphics[width=13cm]{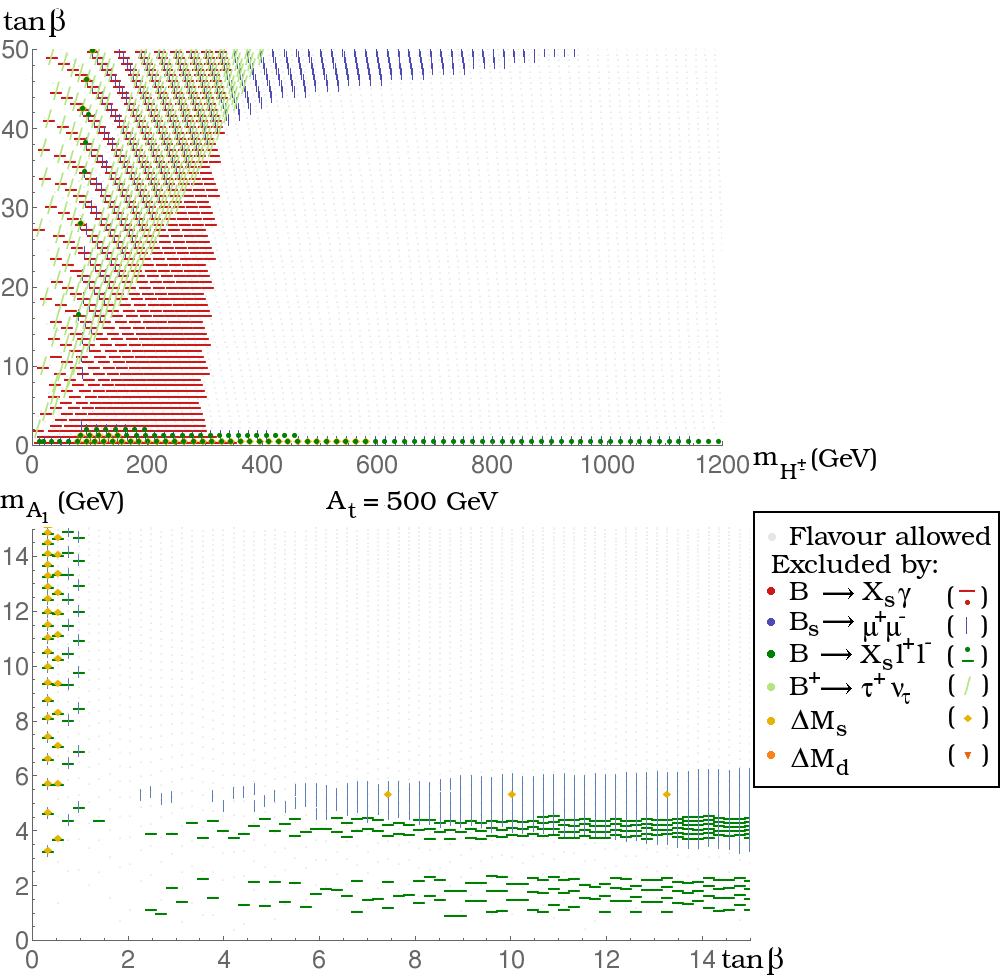}
  \caption{Top: Exclusion contours due to the flavour constraints in the plane $\{m_{H^{\pm}},\tan\beta\}$ for $\lambda=\kappa=2\cdot10^{-4}$, 
$\mu=2M_1=M_2=M_3/3=500$~GeV, $m_{\tilde{F}}=1$~TeV, $A_t=0.5$~TeV, $A_{b,\tau}=-1.5$~TeV, $A_{\kappa}=-500$~GeV. Bottom: Exclusion contours due to the flavour constraints in the plane $\{\tan\beta,m_{A_1}\}$ for $\lambda=0.45$ $\kappa=0.4$, 
$\mu=2M_1=M_2=M_3/3=500$~GeV, $m_{\tilde{F}}=1$~TeV, $A_{b,\tau}=-1.5$~TeV, $A_t=0.5$~TeV, $A_{\kappa}=0$~GeV. Both plots are obtained with the new version of the code. The colour (symbol) code remains the same as before.}
  \label{At500}
\end{figure}
To counteract this effect, we present in Fig.~\ref{At500} the limits from flavour processes obtained with the new implementation, for $A_t=500$~GeV and a somewhat heavier chargino / neutralino sector. The plot on the top 
again considers the plane $\{m_{H^{\pm}},\tan\beta\}$ in the MSSM limit. SUSY contributions are suppressed by the choice of low $A_t$. Correspondingly, limits from $\bar{B}\to X_s\gamma$ only intervene in the region with low 
$m_{H^{\pm}}<300$~GeV. The constraints driven by $B_s^0\to\mu^+\mu^-$ eventually exclude the large $\tan\beta\gsim45$ range but are obviously weaker than before. On the other hand, the exclusion contour associated with 
$B^+\to\tau^+\nu_{\tau}$ and $\bar{B}\to X_sl^+l^-$ remain largely unaffected. The plot on the bottom part of Fig.~\ref{At500} addresses the scenario with a light pseudoscalar: contrarily to the case of Fig.~\ref{MA1-TB},
CP-odd masses above $m_{A_1}\sim6$~GeV are left unconstrained by the flavour test, with exclusions intervening only at very-low $\tan\beta$ or for $m_{A_1}$ in the immediate vicinity of a resonant energy (for $B_s^0\to\mu^+\mu^-$,
$\Delta M_s$ or $\bar{B}\to X_sl^+l^-$).

\subsection{Impact of the new flavour tests}
Beyond the observables which had been considered in \cite{Domingo:2007dx}, we have extended our analysis to several new channels. We now wish to discuss their
impact on the NMSSM parameter space.

\begin{figure}[!htb]
    \centering
    \includegraphics[width=13cm]{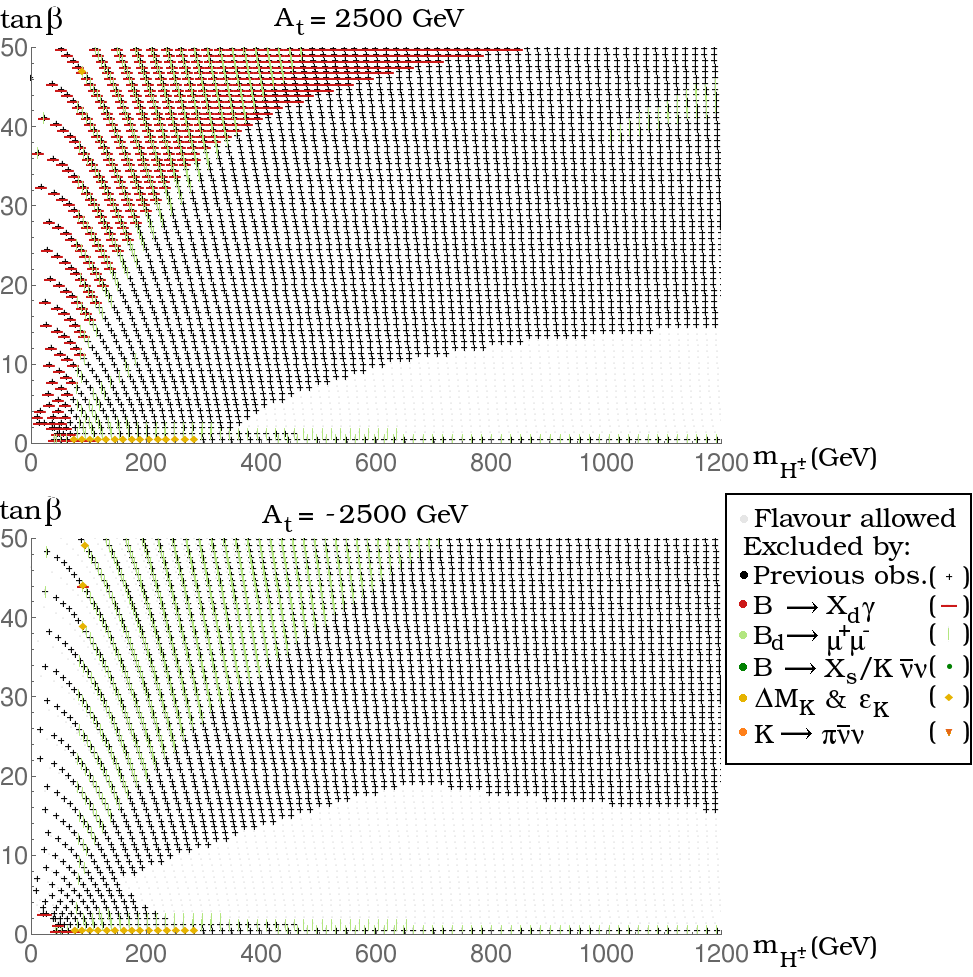}
  \caption{Exclusion contours driven by $\bar{B}\to X_d\gamma$ (red, horizontal lines), $B_d^0\to\mu^+\mu^-$ (light green, vertical lines), $B\to X_s/K\bar{\nu}\nu$ (dark green, circles), the $K-\bar{K}$ mixing (yellow, diamonds) and 
$K\to\pi\bar{\nu}\nu$ (orange, triangle) in the plane $\{m_{H^{\pm}},\tan\beta\}$ for the scenario of Fig.~\ref{MHC-TB}. The limits obtained with the observables considered in Fig.~\ref{MHC-TB} are shown on the background in black
(crosses). The case of $A_t>0$ is depicted on the top, while $A_t<0$ is on the bottom.}
  \label{MHC-TB_NC}
\end{figure}
In Fig.~\ref{MHC-TB_NC}, we consider the scenario of Fig.~\ref{MHC-TB} once more and present the exclusion limits driven by the newly implemented channels. Note that the constraints considered in the previous section form the black 
exclusion zone on the background. The limits from the various channels shown in this plane seem to be essentially subleading in view of these previous constraints of Fig.~\ref{MHC-TB}.
\begin{itemize}
 \item Limits from $\bar{B}\to X_d\gamma$ intervene essentially for $A_t>0$ (i.e.\ for constructive SUSY contributions), large $\tan\beta$ (driving large SUSY contributions) and light $H^{\pm}$ (driving large 2HDM contributions). Yet 
the corresponding bounds are superseded by $\bar{B}\to X_s\gamma$.
 \item $B_d^0\to\mu^+\mu^-$ intervenes in the large $\tan\beta$ / low $m_{H^{\pm}}$ corner as well, e.g.\ for $A_t<0$, but seems less sensitive than $B_s^0\to\mu^+\mu^-$, except in the low $\tan\beta\lsim2$ region.
 \item The processes of the $b\to s\bar{\nu}\nu$ and $s\to d\bar{\nu}\nu$ transitions are found to be well under the current experimental upper bounds.
 \item The $K-\bar{K}$ mixing excludes a few points (driven by $\varepsilon_K$ where the SM is slightly off, with respect to the experimental results) but is not competitive in view of the, admittedly conservative, uncertainties.
\end{itemize}

Note that the limits induced by the $b\to c\tau\nu_{\tau}$ channels have been omitted in Fig.~\ref{MHC-TB_NC}. Given the current data, this transition would exclude the whole $\{m_{H^{\pm}},\tan\beta\}$ plane, with the exception of
the large $\tan\beta$ / low $m_{H^{\pm}}$ corner -- which is excluded by most of the other flavour constraints: the significant discrepancy of the SM estimate with the experimental measurement, especially for $B\to D^*\tau\nu_{\tau}$, 
explains this broad exclusion range. SUSY 2HDM effects cannot reduce the gap much, except in already excluded regions of the parameter space.

\begin{figure}[!htb]
    \centering
    \includegraphics[width=13cm]{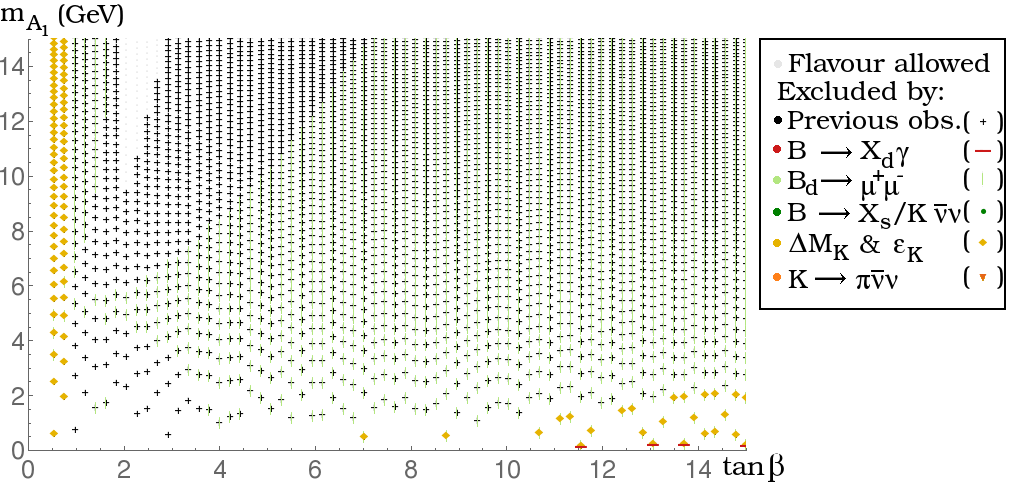}
  \caption{Exclusion contours driven by $\bar{B}\to X_d\gamma$, $B_d^0\to\mu^+\mu^-$, $B\to X_s/K\bar{\nu}\nu$, the $K-\bar{K}$ mixing and $K\to\pi\bar{\nu}\nu$ in the plane $\{\tan\beta,m_{A_1}\}$ for the scenario of 
Fig.~\ref{MA1-TB}. The limits obtained with the observables considered in Fig.~\ref{MA1-TB} are shown on the background in black. We employ the same colour / symbol code as for Fig.~\ref{MHC-TB_NC}.}
  \label{MA1-TB_NC}
\end{figure}
Then, we return to the light pseudoscalar scenario of Fig.~\ref{MA1-TB} and display the constraints associated with the new channels in Fig.~\ref{MA1-TB_NC}. Again, these limits are found to be weaker than those shown in the previous
section. Limits from $B_d^0\to\mu^+\mu^-$ prove the most constraining of the new channels in this regime: this again results from the enhancement of the Higgs-penguin mediated by a resonant $A_1$. Subleading constraints from the 
$K-\bar{K}$ mixing also intervene at low $\tan\beta\lsim1$ and for very light CP-odd Higgs with $m_{A_1}\lsim2$~GeV. Again, the discrepancy among SM predictions and experimental measurements for the $b\to c\tau\nu_{\tau}$ transition 
cannot be interpreted in this scenario, so that applying a $95\%$ CL test for the ratios $R_{D^{(*)}}$ would lead to the exclusion of the whole portion of parameter space displayed in Fig.~\ref{MA1-TB_NC}.

\begin{figure}[!htb]
    \centering
    \includegraphics[width=13cm]{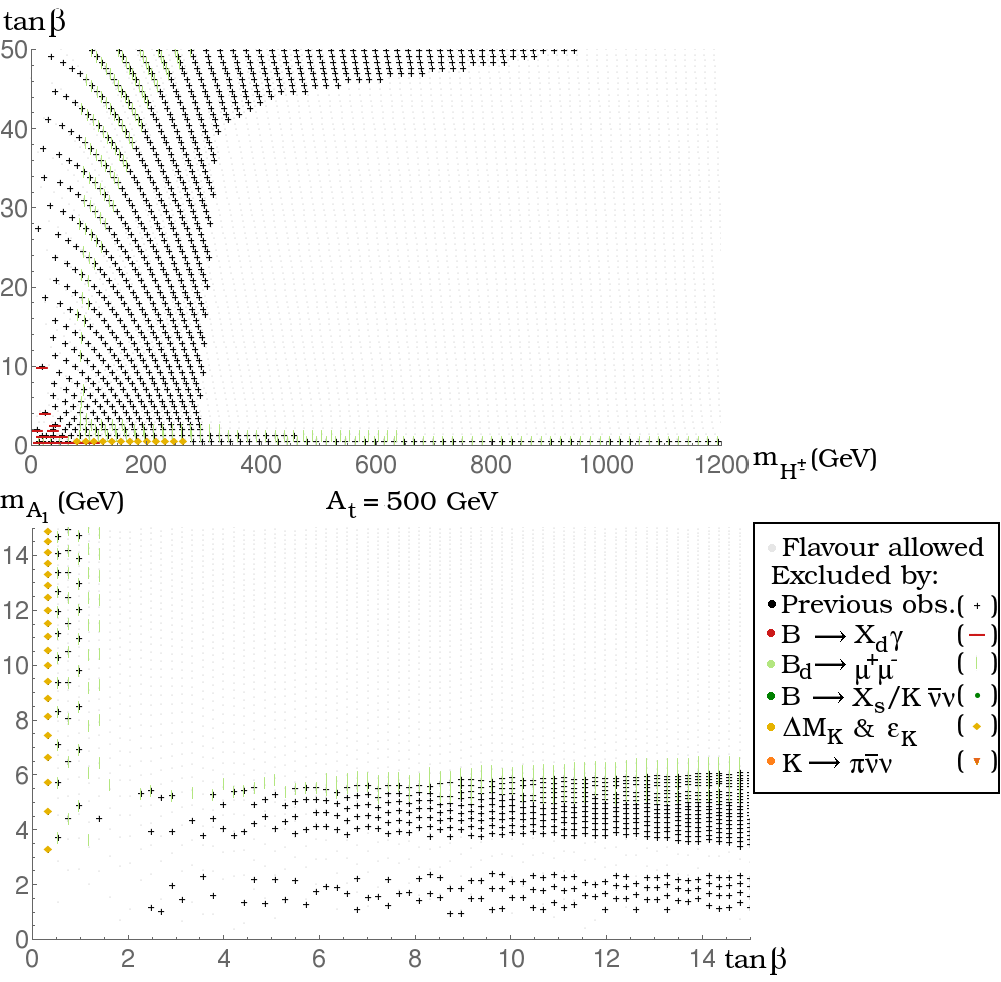}
  \caption{Exclusion contours driven by $\bar{B}\to X_d\gamma$, $B_d^0\to\mu^+\mu^-$, $B\to X_s/K\bar{\nu}\nu$, the $K-\bar{K}$ mixing and $K\to\pi\bar{\nu}\nu$
for the scenarii of Fig.~\ref{At500}. The limits obtained with the observables considered in Fig.~\ref{At500} are shown on the background. The colour / symbol code is left unchanged.}
  \label{At500_NC}
\end{figure}

Finally, we complete this discussion by considering the parameter sets of Fig.~\ref{At500}, where the flavour limits discussed in the previous section were found weaker. The impact of the new channels can be read in Fig.~\ref{At500_NC}
The corresponding exclusion regions in the considered regime with $A_t=500$~GeV again prove narrower than those considered in Fig.~\ref{At500}. (Note again that we have omitted the $b\to c\tau\nu_{\tau}$ channels, however.)

Therefore, we find that the new channels tested in \verb|bsg.f| and \verb|Kphys.f| are typically less constraining than the older ones, which we discussed before. Limits from $\bar{B}\to X_d\gamma$ and $B_d^0\to\mu^+\mu^-$ are 
found to be significant, however, and an evolution of the experimental limits or an improvement in understanding the SM uncertainties may provide them with more relevance in the future. The $b\to c\tau\nu_{\tau}$ transition stands 
apart as the tension between SM and experiment resists an NMSSM interpretation, at least in the scenarios that we have been considering here.

\section{Conclusions}
We have considered a set of flavour observables in the NMSSM, updating and extending our former analysis in \cite{Domingo:2007dx}. These channels have been implemented in a pair of Fortran subroutines, which allow for both the 
evaluation of the observables in the NMSSM and confrontation with the current experimental results. We have taken into account the recent upgrades of the SM status of e.g.\ $BR[\bar{B}\to X_s\gamma]$ or $BR[B_s^0\to\mu^+\mu^-]$
and included BSM effects at NLO. The tools thus designed will be / have been partially made public within the package \verb|NMSSMTools| \cite{NMSSMTools}.

We observe that the bounds on the NMSSM parameter space driven by $BR[\bar{B}\to X_s\gamma]$ or $BR[B_s^0\to\mu^+\mu^-]$ have become more efficient, which should be considered in the light of the recent evolution of the SM status 
and/or the experimental measurement for both these channels. In particular, the large $\tan\beta$ region is rapidly subject to constraints originating from the flavour sector. Similarly, the light pseudoscalar scenario is tightly corseted due to the 
efficiency of Higgs-penguins in the presence of such a light mediator.

Among the new channels that we have included, we note the specific status of the $b\to c\tau\nu_{\tau}$ transition, where the discrepancy between SM and experiment seems difficult to address in a SUSY context.

Other channels of the flavour-changing sector may prove interesting to include in the future. Note e.g.\ the current evolution in the $B\to K^{(*)}l^+l^-$ observables.

\section*{Acknowledgements}
The author is grateful to U.~Ellwanger for useful comments and thanks D.~Barducci for spotting a numerical instability in the pre-released version of \verb|bsg.f|.
This work is supported by CICYT (grant FPA 2013-40715-P).

\end{document}